\RequirePackage{fix-cm}
\documentclass[twocolumn,epjc3]{svjour3}  
\smartqed  
\RequirePackage{graphicx}
\RequirePackage{amsmath}
\RequirePackage{amssymb}
\RequirePackage{wasysym}
\RequirePackage{color}
\RequirePackage{booktabs}
\RequirePackage{lineno}

\definecolor{arsenic}{rgb}{0.23,0.27,0.29}
\definecolor{grey}{rgb}{0.6,0.6,0.6}
\definecolor{darkgrey}{rgb}{0.4,0.4,0.4}
\definecolor{darkblue}{RGB}{0,0,127}
\definecolor{darkgreen}{RGB}{0,170,0}
\definecolor{darkred}{RGB}{220,0,0}

\newcommand{\onbb}{$0\nu\beta\beta$}
\newcommand{\meff}{$|m_{\beta\beta}|$}
\newcommand{\mmin}{$m_{min}$}
\newcommand{\mbeta}{$m_{\beta}$}
\newcommand{\Tonbb}{$T_{1/2}^{0\nu}$}

\newcommand{\Qbb}{$Q_{\beta\beta}$}
\newcommand{\bb}{$\beta\beta$}
\newcommand{\Ge}{$^{76}$Ge}
\newcommand{\Te}{$^{130}$Te}
\newcommand{\Xe}{$^{136}$Xe}

{
\journalname{Eur. Phys. J. C}
\begin{document}

\title{Effective Majorana Mass and Neutrinoless Double Beta Decay
}


\author{Giovanni Benato\thanksref{e1,addr1}}

\thankstext{e1}{e-mail: gbenato@physik.uzh.ch}


\institute{Physik Institut der Universit\"at Z\"urich, Zurich, Switzerland \label{addr1}}

\date{Received: date / Accepted: date}

\maketitle

\begin{abstract}
The probability distribution for the  effective Majorana mass
as a function of the lightest neutrino mass in the standard three neutrino scheme
is computed via a random sampling from the distributions of the involved
mixing angles and squared mass differences.
A flat distribution in the $[0,2\pi]$ range for the Majorana phases is assumed,
and the dependence of small values of the effective mass on the Majorana phases is highlighted.
The study is then extended with the addition of the cosmological bound
on the sum of the neutrino masses.
Finally, the prospects for \onbb\ decay search with \Ge, \Te\ and \Xe\ are discussed,
as well as those for the measurement of the electron neutrino mass.
\end{abstract}

\section{Introduction}
\label{intro}

Neutrinoless double beta decay (\onbb) is a topic of major interest for the present
and near future of neutrino physics~\cite{Bilenky:2014uka}.
Its observation would prove the violation
of total lepton number conservation and  provide information on the absolute neutrino mass scale.
In the following the assumption is made that the standard light neutrino exchange
is the dominant contribution.
The parameter of interest in \onbb\ decay is the so-called effective Majorana mass, \meff,
which depends on the three neutrino mass eigenstates, on two of the PMNS mixing angles,
and on two Majorana phases~\cite{Bilenky:2014uka}.
Currently, a good knowledge of the two squared mass differences
and of the mixing angles is available thanks to accelerator and reactor experiments~\cite{Capozzi:2013csa}.
Moreover, some bound on the sum of neutrino masses is available
from cosmological observations~\cite{Ade:2015xua,Palanque-Delabrouille:2014jca}.
On the contrary, assuming that neutrinos are Majorana particles,
no information is available on the Majorana phases.

Typically, the allowed range for \meff\ as a function
of the lightest neutrino mass is calculated via an error propagation
on the involved mixing angles and squared mass differences,
with no constraint on the Majorana phases.
Recently, the cosmological limit has been introduced, too~\cite{Dell'Oro:2015tia}.

In the present paper, an alternative approach for the extraction
of the effective mass allowed regions is exploited.
Instead of presenting only e.g. $3~\sigma$ coverage regions for the \meff,
a probability distribution is given, which can be valuable for the design of experiments.
This is extracted via random sampling on the probability distributions of the measured parameters
and on a flat distribution for the Majorana phases.

After a short formulation of the problem in Sec.~\ref{sec:EffectiveMass},
the probability distribution for \meff\ as a function of the lightest neutrino mass
is reported in Sec.~\ref{sec:EffMassVsMmin}.
The \meff\ dependence on the values of the Majorana phases is highlighted,
and the case of vanishing \meff\ is discussed.
In Sec.~\ref{sec:EffMassCosmology}, the cosmological bound on the sum of the neutrino masses
in inserted in the calculation and its influence in the distributions is described.
The perspectives for \onbb\ decay search given this study of \meff\
and the present knowledge of the nuclear matrix elements are presented in Sec.~\ref{sec:halflife}.
Additionally, the probability distribution for \meff\ as a function
of the electron neutrino mass is given in Sec.~\ref{sec:mbeta},
and the physics reach of beta spectrum end point measurements is discussed.

The aim of this paper is to prove the possibility of using the information on the sum
of neutrino masses provided by the cosmological measurements,
and to demonstrate the dependence of the effective mass
on the assumption made for the Majorana phases.
The comparison of such different assumptions can provide
a deeper understanding of the current status and the perspective of \onbb\ decay search.

\section{Effective Majorana Mass}
\label{sec:EffectiveMass}

The parameter of interest in the \onbb\ decay search is \meff.
It is a combination of the neutrino mass eigenstates and the neutrino mixing matrix terms.
Under the hypothesis that only the known three light neutrinos participate in the process,
the effective mass is given by:
\begin{equation}\label{eq:effMass}
  | m_{\beta\beta}| = \Biggl| \sum_{i=1}^3 U_{ei}^2 m_i \Biggr|
\end{equation}
where $U$ is the PMNS mixing matrix~\cite{Bilenky:2014uka}, with two additional Majorana phases.
The expansion of Eq.~\ref{eq:effMass} yields:
\begin{equation}\label{eq:effMass2}
| m_{\beta\beta} | = \bigl|  c_{12}^2c_{13}^2 m_1 + s_{12}^2 c_{13}^2 m_2 e^{i\alpha} + s_{13}^2 m_3 e^{i\beta}  \bigr|
\end{equation}
where the CP-violating phase present in the PMNS matrix
is hidden in the Majorana phases $\alpha$ and $\beta$.
In this formulation, the symbols $c_{jk}(s_{jk})$ stay for $\cos{\theta_{jk}}(\sin{\theta_{jk}})$.
Expanding Eq.~\ref{eq:effMass2} and following the definition of absolute value
for complex numbers:
\begin{multline}\label{eq:effMass3}
  | m_{\beta\beta} | = \sqrt{ \bigl( c_{12}^2 c_{13}^2 m_1 + s_{12}^2 c_{13}^2 m_2 \cos{\alpha} + s_{13}^2 m_3 \cos{\beta} \bigr)^2 +}\\
    \overline{ + \bigl( s_{12}^2 c_{13}^2 m_2 \sin{\alpha} + s_{13}^2 m_3 \sin{\beta} \bigr)^2 }
\end{multline}
The parameters involved are:
\begin{itemize}
\item the angles $\theta_{12}$ and $\theta_{13}$, measured with good precision
  by the solar and short-baseline reactor neutrino experiments, respectively;
\item the neutrino mass eigenstates $m_1$, $m_2$ and $m_3$,
  which are related to the solar and atmospheric squared mass differences
  $\delta m_{\sun}^2$ and $\Delta m_{atm}^2$:
  \begin{align}\label{eq:deltaMs}
    \delta m_{\sun}^2 & \simeq \Delta m_{21}^2 \nonumber \\
    \Delta m_{atm}^2 & \simeq \frac{1}{2} \bigl| \Delta m_{31}^2 + \Delta m_{32}^2 \bigr|
  \end{align}
  These are known with $\sim3\%$ uncertainty thanks to long-baseline reactor
  and long-baseline accelerator neutrino experiments, respectively (see Tab.~\ref{tab:oscillationparameters}).
  The mass eigenstates are also related to the sum of neutrino masses:
  \begin{equation}\label{eq:sum}
    \Sigma = \sum_{i=1}^3 m_i
  \end{equation}
  for which several upper limits of about 0.1-0.2~eV are set
  by cosmological observations~\cite{Ade:2015xua,Palanque-Delabrouille:2014jca,Dell'Oro:2015tia};
\item the two Majorana phases $\alpha$ and $\beta$,
  for which no experimental information is available.
\end{itemize}

The relation between the mass eigenstates and the squared mass differences
given in Eq.~\ref{eq:deltaMs} allows two possible orderings of the neutrino masses~\cite{Agashe:2014kda}.
Using the same notation of~\cite{Giunti:2015kza}, a first scheme, denoted as Normal Hierarchy (NH), corresponds to:
\begin{align}\label{eq:NH}
  m_1 & = m_{min} \nonumber \\
  m_2 & = \sqrt{ m_{min}^2 + \delta m_{\sun}^2 } \nonumber \\
  m_3 & = \sqrt{ m_{min}^2 + \Delta m_{atm}^2 + \frac{\delta m_{\sun}^2}{2} }
\end{align}
where \mmin\ is the mass of the lightest neutrino. The so-called Inverted Hierarchy (IH) is given by:
\begin{align}\label{eq:IH}
  m_1 & = \sqrt{ m_{min}^2 + \Delta m_{atm}^2 - \frac{\delta m_{\sun}^2}{2} } \nonumber \\
  m_2 & = \sqrt{ m_{min}^2 + \Delta m_{atm}^2 + \frac{\delta m_{\sun}^2}{2} } \nonumber \\
  m_3 & = m_{min}
\end{align}
Present data do not show any clear preference for either of the two schemes.

\section{Effective Mass Versus Lightest Neutrino Mass}
\label{sec:EffMassVsMmin}

The effective mass can be expressed  as a function of the lightest neutrino mass,
as first introduced in~\cite{Vissani:1999tu}.
This is normally done via a $\chi^2$ analysis~\cite{Giunti:2015kza},
where the uncertainties on the mixing angles $\theta_{12}$ and $\theta_{13}$,
and on the squared mass differences $\delta m_{\sun}^2$ and $\Delta m_{atm}^2$
are propagated, while the values of the Majorana phases leading to the largest
and smallest \meff\ are considered.
As a result, the $1\sigma$, $2\sigma$ and $3\sigma$ allowed regions are typically shown (e.g. Fig.~3 of~\cite{Giunti:2015kza}),
but no clear information about the relative probability
of different \meff\ values for a fixed \mmin\ is provided.
This can become of dramatic importance in case future experiments
prove that nature chose the NH regime.
In NH, the effective mass is distributed within a flat area between $\sim10^{\mbox{-}3}$ and $\sim5\cdot10^{\mbox{-}3}$~eV
if $m_{min}<10^{\mbox{-}3}$~eV, while it can vanish for $m_{min} \in [10^{\mbox{-}3},10^{\mbox{-}2}]$~eV
due to the combination of the Majorana phases.
The case of a vanishing \meff\ is possible only
for a smaller subset of values of the Majorana phases than the whole $[0,2\pi]$ range.
This does not necessarily mean that a vanishing effective mass
implies that the theory suffers of dangerous fine tuning.
Namely, in some models the effective mass can assume a naturally small value 
that remains small after renormalization
due to the chiral symmetry of fermions~\cite{Vissani:2003aj,Vissani:2001im}.

Without giving a preference to any model,
one can ask which is the distribution probability of \meff\
for a fixed value of \mmin\ and, moreover, which is the probability for the NH case
of having $|m_{\beta\beta}| < 10^{\mbox{-}3}$~eV
given the present knowledge (or ignorance) of the various parameters involved.
An answer is obtained using a toy Monte Carlo (MC) approach,
where a random number is sampled for each parameter according to its (un)known measured value,
and \meff\ is computed for each trial.

\begin{table}
  \caption{\label{tab:oscillationparameters}Parameters for the evaluation
    of the effective Majorana mass.}
  \begin{tabular}{ll}
    \hline\noalign{\smallskip}
    Parameter & Value~\cite{Capozzi:2013csa} \\
    \noalign{\smallskip}\hline\noalign{\smallskip}
    $\delta m_{\sun}^2$      & $(7.54\pm0.26)\cdot10^{\mbox{-}5}$~eV$^2$ \\
    $\Delta m_{atm}^2$ (NH) & $(2.43\pm0.06)\cdot10^{\mbox{-}3}$~eV$^2$ \\
    $\Delta m_{atm}^2$ (IH) & $(2.38\pm0.06)\cdot10^{\mbox{-}3}$~eV$^2$ \\
    $s_{12}^2$              & $(3.08\pm0.17)\cdot10^{\mbox{-}1}$ \\
    $s_{13}^2$ (NH)         & $(2.34\pm0.20)\cdot10^{\mbox{-}2}$ \\
    $s_{13}^2$ (IH)         & $(2.40\pm0.22)\cdot10^{\mbox{-}2}$ \\
    \noalign{\smallskip}\hline\noalign{\smallskip}
    $\Sigma$               & $(22\pm62)\cdot10^{\mbox{-}3}$~eV~\cite{Palanque-Delabrouille:2014jca,Dell'Oro:2015tia} \\
    \noalign{\smallskip}\hline
  \end{tabular}
\end{table}

The values for the experimentally measured parameters are taken from Tab.~3 of~\cite{Capozzi:2013csa}.
In case the upper and lower error on some parameter are different,
the random sampling is performed
using a Gaussian distribution with mean given by the best fit of~\cite{Capozzi:2013csa}
and $\sigma$ given by the greater among the upper and lower uncertainties.
These values are reported in Tab.~\ref{tab:oscillationparameters}.
The effect of this conservative choice on the resulting allowed regions for \meff\ vs \mmin\ is small,
and the message of this study is not changed.
Similarly, the use of a more recent and precise value for $\theta_{13}$~\cite{An:2015rpe}
does not significantly affect the result.
An eventual correlation between the involved parameters could be easily included
in the study. For the values of Tab.~\ref{tab:oscillationparameters}
it can be considered negligible and is not taken into account.

The choice not to prefer any model is reflected on the distribution assigned to the Majorana phases.
Assuming a complete ignorance on $\alpha$ and $\beta$,
their values are sampled from a flat distribution in the $[0,2\pi]$ range.

\begin{figure}
  \centering
  \def\svgwidth{\columnwidth}
  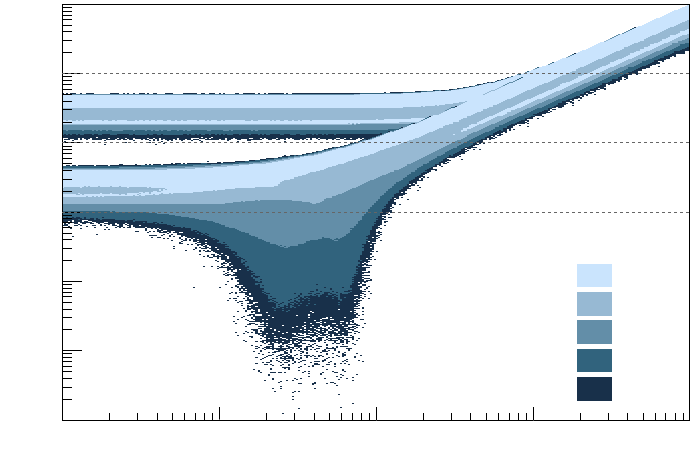
  \caption{Effective Majorana mass as a function of the lightest neutrino mass.
    The top band correspond to the IH regime, the bottom to NH.
    The different colors correspond to the 1,\dots,5~$\sigma$ coverage regions.}
  \label{fig:NoC_Meff_Mmin}
\end{figure}

In order to keep the bi-logarithmic scale normally used in literature
but with the aim of maintaining the same normalization over all the considered area,
a two dimensional histogram with increasing bin size is exploited.
In particular, the bin width $\Delta_i$ is given by $\Delta_i = k \Delta_{i\mbox{-}1}$ with $k>1$,
for both the $x$ and the $y$ directions.
For each bin of the $x$-axis, $10^6$ random parameter combinations are used to calculate
the probability distribution for \meff, leading to the plot shown in Fig.~\ref{fig:NoC_Meff_Mmin}.
The different color levels correspond to the 1, 2, \dots, 5 $\sigma$ coverage regions.
The sensitivity of current experiments, at the $10^{\mbox{-}1}$~eV level, is reported,
together with the sensitivity of an hypothetical ton scale experiment
and the ultimate sensitivity of a 100 ton scale setup with the assumption of zero background.

\begin{figure}
  \centering
  \def\svgwidth{\columnwidth}
  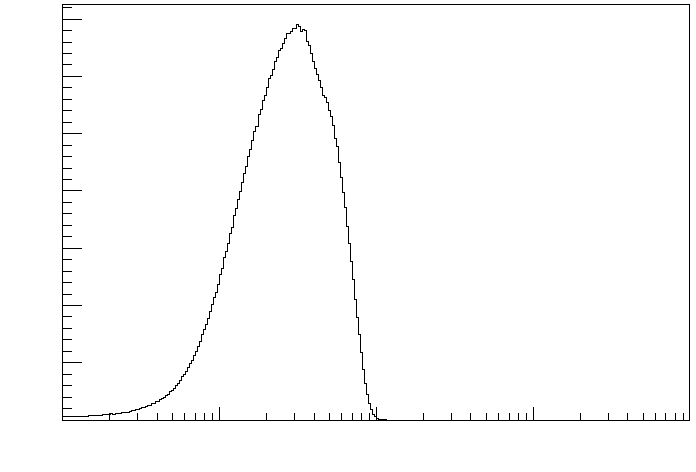
  \caption{Probability for $|m_{\beta\beta}|<10^{\mbox{-}3}$~eV in the NH regime.}
  \label{fig:NoC_MeffProb}
\end{figure}
\begin{figure}
  \centering
  \def\svgwidth{\columnwidth}
  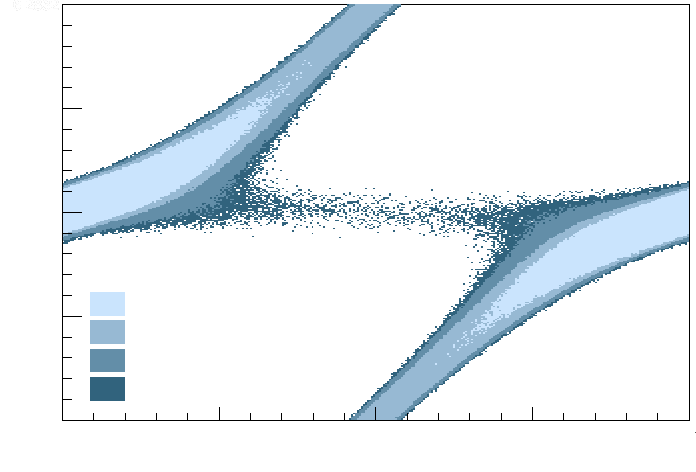
  \caption{Majorana phases $\alpha$ and $\beta$ for $|m_{\beta\beta}|<10^{\mbox{-}3}$~eV in the NH regime.
    The different colors correspond to the 1,\dots,4~$\sigma$ coverage regions.}
  \label{fig:NoC_phases_smallMeff}
\end{figure}

Looking at the \meff\ population for both the IH and NH,
high values of \meff\ are favored for all values of \mmin.
This can have a strong impact on the perspectives of \onbb\ decay search in the next decades.
For the NH case, the probability of having $|m_{\beta\beta}| < 10^{\mbox{-}3}$~eV
is reported in Fig.~\ref{fig:NoC_MeffProb}. Even for the most unfortunate case
of $m_{min}\sim3\mbox{--}4\cdot10^{\mbox{-}3}$~eV, given the present knowledge of the oscillation parameters
there is at least $93\%$ probability of detecting a \onbb\ decay signal if an experiment
with $10^{\mbox{-}3}$~eV discovery sensitivity on \meff\ is available.
Such a sensitivity would involve the realization of an experiment with  $\sim100$~ton active mass
operating in zero-background condition (see Sec.~\ref{sec:halflife}).
If this is presently hard to imagine, some case studies are already published on the topic~\cite{Biller:2013wua}.
On the other side, the creation of an experiment with $10^{\mbox{-}5}$~eV sensitivity
would most probably be out of reach because it would involve the deployment
of $\gtrsim10^6$~ton of active material.

One can ask which values of the Majorana phases are needed in order to obtain
$|m_{\beta\beta}|<10^{\mbox{-}3}$~eV. This is shown in Fig.~\ref{fig:NoC_phases_smallMeff}:
small values of the effective mass are only possible if $\alpha$ and $\beta$
differ by a value $\sim\pi$.
With reference to Eq.~\ref{eq:effMass3}, neglecting for the moment the term $c_{12}^2c_{13}^2m_1$
and supposing all other terms have the same amplitude,
\meff\ approaches zero only if both couples ($\sin{\alpha}$, $\sin{\beta}$) 
and ($\cos{\alpha}$, $\cos{\beta}$) have opposite signs.
The condition is satisfied only if $\alpha$ and $\beta$ belong to opposite quadrants.
Considering the amplitude of the terms, the major difference is that
\meff\ can become small for $m_{min}\in[10^{\mbox{-}3},10^{\mbox{-}2}]$~eV and not  for $m_{min}=0$,
but the required correlation between the Majorana phases is unchanged.
Hence, our result shows that in the type of models~\cite{Vissani:2003aj,Vissani:2001im} mentioned above
the Majorana phases are closely correlated.

One remark has to be made regarding the sparsely populated region for $\alpha\in[\pi/2,3\pi/2]$ and $\beta\sim\pi$
of Fig.~\ref{fig:NoC_phases_smallMeff}. These points correspond to those in the region
with $|m_{min}|<10^{\mbox{-}3}$~eV and $|m_{\beta\beta}|<10^{\mbox{-}3}$~eV of Fig.~\ref{fig:NoC_Meff_Mmin},
or in other words to the bottom left part of the horizontal NH band.
Hence, they can be considered a spurious contamination coming
from the choice of selecting the events with $|m_{min}|<10^{\mbox{-}3}$~eV.

\section{How Does Cosmology Affect $0\nu\beta\beta$ Decay Search?}
\label{sec:EffMassCosmology}

In the analysis presented so far the effective mass depends on
the three free parameters: the two Majorana phases
are considered as nuisance parameters with uniform distribution,
and the probability distribution for \meff\ as a function of \mmin\ is obtained.

Several cosmological measurements allow to put upper bounds on the sum of neutrino masses, $\Sigma$.
These limits are typically around $0.1$--$0.2$~eV~\cite{Ade:2015xua,Palanque-Delabrouille:2014jca,Dell'Oro:2015tia},
depending on the considered data sets.
Recently, a combined analysis of the Planck 2013 data and several Lyman-$\alpha$ forest data sets
lead to a Gaussian probability distribution for $\Sigma$,
with $\Sigma = (22\pm62)\cdot10^{\mbox{-}3}$~eV~\cite{Palanque-Delabrouille:2014jca,Dell'Oro:2015tia}.
The corresponding $95\%$~CL limit is $\Sigma<0.146$~eV.
The distribution was already used in~\cite{Dell'Oro:2015tia} to extract the allowed range
for \meff\ as a function of $\Sigma$.
In that case, the allowed regions for NH and IH are weighted with the cosmological bound on $\Sigma$,
and it is pointed out that the allowed region for IH is strongly reduced.

The study can be extended including the cosmological bound
following a different approach than that used in~\cite{Dell'Oro:2015tia}.
Considering Eqs.~\ref{eq:NH}, \ref{eq:IH} and~\ref{eq:sum},
the three parameters $m_1$, $m_2$ and $m_3$ depend on $\delta m_{\sun}^2$,
$\Delta m_{atm}^2$ and $\Sigma$, for which a measurement is available.
Hence, a random sampling is performed on $\delta m_{\sun}^2$, $\Delta m_{atm}^2$ and $\Sigma$,
and the values of the mass eigenstates are extracted numerically
after solving the system of Eqs.~\ref{eq:NH}, \ref{eq:sum} for NH,
and~\ref{eq:IH}, \ref{eq:sum} for IH.
Considering the NH case and given the measured values of the squared mass differences,
Eq.~\ref{eq:NH} states that the minimum value of $\Sigma$ is:
\begin{equation}\label{eq:sigmaMinNH}
  \Sigma_{min}^{NH} = \sqrt{ \delta m_{\sun}^2 } + \sqrt{ \Delta m_{atm}^2 + \frac{\delta m_{\sun}^2}{2} }
  \simeq 0.058~\text{eV}
\end{equation}
where \mmin\ has been set to zero.
Similarly, for NH:
\begin{align}\label{eq:sigmaMinIH}
  \Sigma_{min}^{IH} & = \sqrt{ \Delta m_{atm}^2 - \frac{\delta m_{\sun}^2}{2} } + \sqrt{ \Delta m_{atm}^2 + \frac{\delta m_{\sun}^2}{2} } \nonumber \\
  & \simeq 0.098~\text{eV}
\end{align}
The combination of the cosmological bound with the measurements of 
$\delta m_{\sun}^2$ and $\Delta m_{atm}^2$ will therefore induce a probability distribution
for $\Sigma$ with a sharp rise at about 0.058(0.098)~eV and a long high-energy tail
for the NH(IH) regime.

\begin{figure}
  \centering
  \def\svgwidth{\columnwidth}
  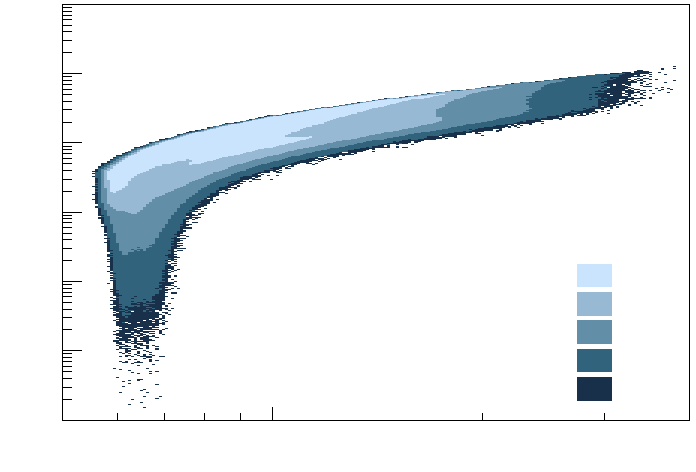
  \caption{Effective Majorana mass as a function of the sum of neutrino masses
    for the NH regime with the application of the cosmological bound.
    The different colors correspond to the 1,\dots,5~$\sigma$ coverage regions.}
  \label{fig:WithC_Meff_sum_NH}
\end{figure}

\begin{figure}
  \centering
  \def\svgwidth{\columnwidth}
  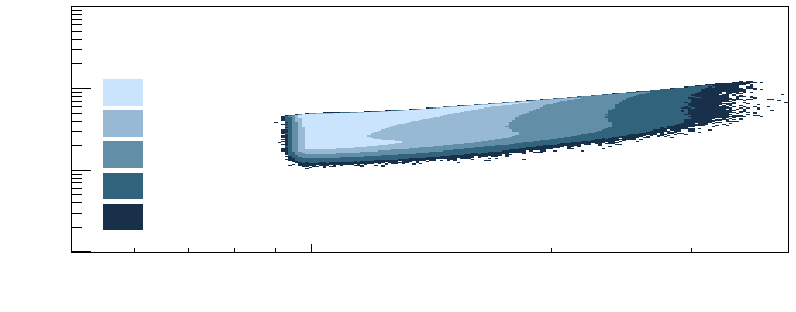
  \caption{Effective Majorana mass as a function of the sum of neutrino masses
    for the IH regime with the application of the cosmological bound.
    The different colors correspond to the 1,\dots,5~$\sigma$ coverage regions.}
  \label{fig:WithC_Meff_sum_IH}
\end{figure}

The probability distributions for \meff\ as a function of $\Sigma$
in the NH and IH cases are shown in Fig.~\ref{fig:WithC_Meff_sum_NH}
and~\ref{fig:WithC_Meff_sum_IH}, respectively.
In total, $10^8$ points are sampled.
The thresholds on $\Sigma$ correspond to the lower bounds mentioned above,
while the horizontal shading for $\Sigma>10^{\mbox{-}1}$~eV comes from the cosmological bound.
In NH case, the vertical shading for $\Sigma\in[6,7]\cdot10^{\mbox{-}2}$~eV
and $|m_{\beta\beta}|<10^{\mbox{-}3}$~eV is related to the combination
of the Majorana phases, as explained in Sec.~\ref{sec:EffMassVsMmin}.

It is worth mentioning that the approach used here does not allow
to make any statement regarding the overall probability
of the NH with respect to the IH regime.
The plots are populated by generating random numbers for
$\delta m_{\sun}^2$, $\Delta m_{atm}^2$ and $\Sigma$.
If $\Sigma > \Sigma_{min}$, with $\Sigma_{min}$ given by Eqs.~\ref{eq:sigmaMinNH}
and~\ref{eq:sigmaMinIH}, the three values are accepted and \meff\ is computed,
otherwise another random number is extracted for $\Sigma$
until the condition is satisfied.
In his way, Figs.~\ref{fig:WithC_Meff_sum_NH} and~\ref{fig:WithC_Meff_sum_IH}
are equally populated, and give no hint about the probability
that nature chose either of the two regimes.

\begin{figure}
  \centering
  \def\svgwidth{\columnwidth}
  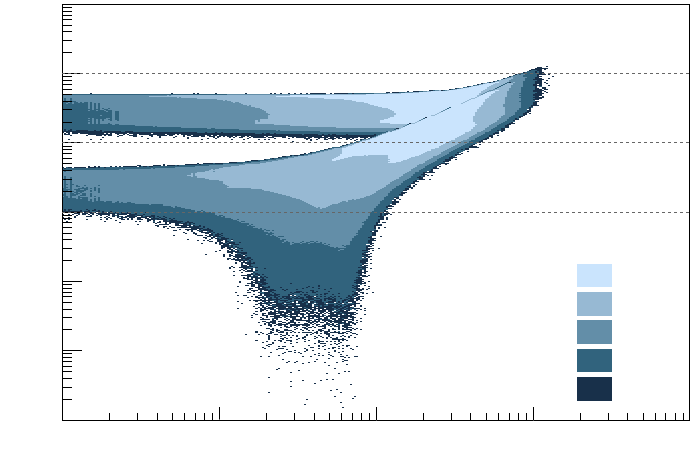
  \caption{Effective Majorana mass as a function of the lightest neutrino mass
    with the application of the cosmological bound.
    The different colors correspond to the 1,\dots,5~$\sigma$ coverage regions.}
  \label{fig:WithC_Meff_Mmin}
\end{figure}

The plot of \meff\ as a function of \mmin\ with the application
of the cosmological bound is shown in Fig.~\ref{fig:WithC_Meff_Mmin}.
Differently from Fig.~\ref{fig:NoC_Meff_Mmin}, it not only provides
the probability distribution for \meff\ as a function of \mmin,
but also the 2-dimensional probability distribution for both parameters together.
The main differences with respect to Fig.~\ref{fig:NoC_Meff_Mmin}
are the decreases of the distribution for $m_{min}\in[5\cdot10^{\mbox{-}2},10^{\mbox{-}1}]$~eV,
and for $m_{min}\lesssim7\cdot10^{\mbox{-}3}$~eV.
Both effects are due to the cosmological bound on $\Sigma$.
Choosing arbitrarily different values for both the mean value
and width for the Gaussian distribution of the total neutrino mass
leads to different shadings on both sides,
and induces the highly populated region around $m_{min}\sim3\cdot10^{\mbox{-}2}$~eV to move.
In particular, a looser bound on $\sigma$ would favor the degenerate mass region,
while a tighter limit would favor smaller values of \mmin, as expected,
with strong consequences for the predicted \meff.
With the present assumptions, values of \meff\ close to the degenerate region are favored.

\begin{figure}
  \centering
  \def\svgwidth{\columnwidth}
  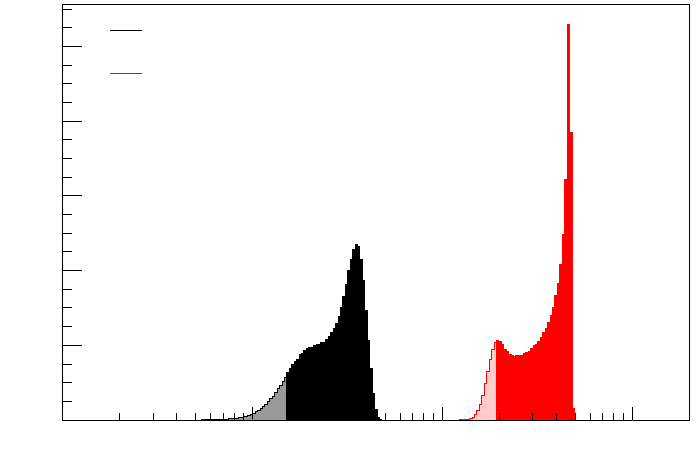
  \caption{Probability distribution for $|m_{\beta\beta}|$ with $m_{min}\in[10^{\mbox{-}4},10^{\mbox{-}3}]$~eV.
    The darker regions correspond to the $90\%$ coverage on \meff.}
  \label{fig:WithC_MeffProb_MminLow}
  \ \\
  \def\svgwidth{\columnwidth}
  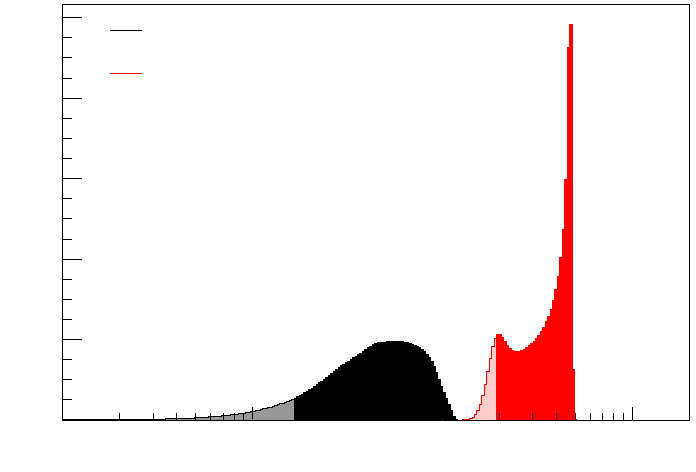
  \caption{Probability distribution for $|m_{\beta\beta}|$ with $m_{min}\in[10^{\mbox{-}3},10^{\mbox{-}2}]$~eV.
    The darker regions correspond to the $90\%$ coverage on \meff.}
  \label{fig:WithC_MeffProb_MminMiddle}
  \ \\
  \def\svgwidth{\columnwidth}
  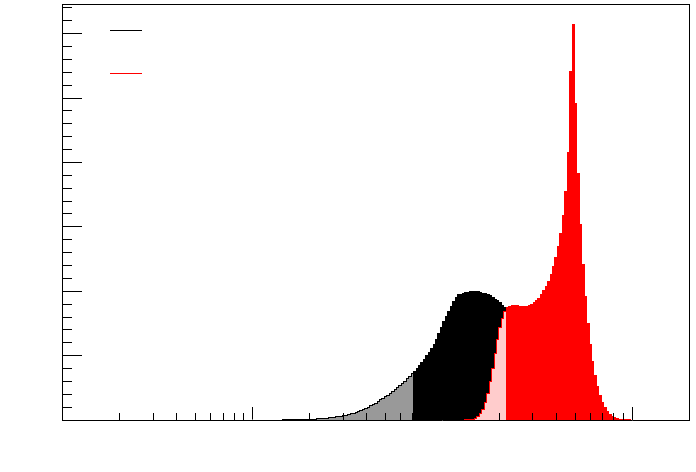
  \caption{Probability distribution for $|m_{\beta\beta}|$ with $m_{min}\in[10^{\mbox{-}2},10^{\mbox{-}1}]$~eV.
    The darker regions correspond to the $90\%$ coverage on \meff.}
  \label{fig:WithC_MeffProb_MminHigh}
\end{figure}

With an eye on the future experiments,
the probability distribution for \meff\ only can be obtained by marginalizing 
the 2-dimensional distribution of Fig.~\ref{fig:WithC_Meff_Mmin} over \mmin.
This is performed separately for three ranges of the lightest neutrino mass:
$m_{min}\in[10^{\mbox{-}4},10^{\mbox{-}3}]$, $m_{min}\in[10^{\mbox{-}3},10^{\mbox{-}2}]$
and $m_{min}\in[10^{\mbox{-}2},10^{\mbox{-}1}]$, as shown in Figs.~\ref{fig:WithC_MeffProb_MminLow},
\ref{fig:WithC_MeffProb_MminMiddle} and~~\ref{fig:WithC_MeffProb_MminHigh}.
In case of a small \mmin\ (Fig.~\ref{fig:WithC_MeffProb_MminLow})
the $90\%$ coverage is obtained for $|m_{\beta\beta}| > 1.54\cdot10^{\mbox{-}3}$~eV and
$|m_{\beta\beta}| > 1.96\cdot10^{\mbox{-}2}$~eV for NH and IH, respectively.
In general, for IH high values of \meff\ are favored.
The $90\%$ coverage on \meff\ for the three considered ranges are reported
in Tab.~\ref{tab:WithC_MeffProb}, together with that
of the overall range $m_{min}\in[10^{\mbox{-}4},1]$~eV.
This case shows how a $3.32\cdot10^{\mbox{-}3}$~eV discovery sensitivity
is required for future experiments in order to have $90\%$ probability
to measure a \onbb\ decay signal in case of NH,
or $2.14\cdot10^{\mbox{-}2}$~eV for IH.

\begin{table}
  \caption{\label{tab:WithC_MeffProb}$90\%$ coverage on \meff\ for NH and IH
    and different ranges of \mmin.}
  \begin{tabular}{ccc}
    \hline\noalign{\smallskip}
          & \multicolumn{2}{c}{\meff\ $90\%$ coverage [eV]}\\
    \mmin~[eV] & NH & IH \\
    \noalign{\smallskip}\hline\noalign{\smallskip}
    $[10^{\mbox{-}4},10^{\mbox{-}3}]$ & $>1.54\cdot10^{\mbox{-}3}$ & $>1.96\cdot10^{\mbox{-}2}$ \\
    $[10^{\mbox{-}3},10^{\mbox{-}2}]$ & $>1.68\cdot10^{\mbox{-}3}$ & $>1.96\cdot10^{\mbox{-}2}$ \\
    $[10^{\mbox{-}2},10^{\mbox{-}1}]$ & $>7.16\cdot10^{\mbox{-}3}$ & $>2.20\cdot10^{\mbox{-}2}$ \\
    \noalign{\smallskip}\hline\noalign{\smallskip}
    $[10^{\mbox{-}4},1]$           & $>3.32\cdot10^{\mbox{-}3}$ & $>2.14\cdot10^{\mbox{-}2}$ \\
    \noalign{\smallskip}\hline
  \end{tabular}
\end{table}

The strong dependence of the result on both the choice
of the Majorana phases distribution and on the cosmological bound
should invoke some caution in the interpretation 
of Figs.~\ref{fig:WithC_Meff_Mmin}, \ref{fig:WithC_MeffProb_MminLow},
\ref{fig:WithC_MeffProb_MminMiddle} and~\ref{fig:WithC_MeffProb_MminHigh}
as the correct probability distributions for \meff.
It is rather important to enlighten the fact that it is in principle possible,
making some arbitrary assumption on the Majorana phases
and provided a reliable cosmological limit on the total neutrino mass,
to extract a probability distribution for \meff.
A higher precision of the cosmological measurement, e.g. by EUCLID~\cite{Laureijs:2011gra,Hamann:2012fe},
would strongly improve the reliability of a prediction on \meff.

\section{From Effective Mass to $0\nu\beta\beta$ Decay Half Life}
\label{sec:halflife}

From an experimental point of view, it is useful to compute the probability distribution
for the \onbb\ decay half life (\Tonbb) as a function of the lightest neutrino mass.
The relation between \Tonbb\ and the effective mass is~\cite{Bilenky:2014uka}:
\begin{equation}\label{eq:T12EffMass}
  \frac{1}{T_{1/2}^{0\nu}} = G^{0\nu} g_A^4 \bigl| {\cal{M}}^{0\nu} \bigr|^2 \frac{|m_{\beta\beta}|^2}{m_e^2}
\end{equation}
where $G^{0\nu}$ is the phase space integral,
$g_A$ the axial vector coupling constant, $\bigl| {\cal{M}}^{0\nu} \bigr|$ the nuclear matrix element (NME),
and $m_e$ the electron mass.
Considering different \bb\ decaying isotopes
and for a fixed effective mass, greater values of the phase space integral and
NME correspond to a shorter \onbb\ decay half life
and, consequently, to a greater specific activity.

If a higher \Qbb\ would in principle allow for a higher phase space integral,
the Coulomb potential of the daughter nucleus plays a strong role in the calculation~\cite{Kotila:2012zza}.
The value of $G^{0\nu}$ is around $10^{\mbox{-}15}$--$10^{\mbox{-}14}$~yr$^{\mbox{-}1}$,
depending on the isotope.

The calculation of $\bigl| {\cal{M}}^{0\nu} \bigr|$ is usually the bottleneck
for the extraction of $|m_{\beta\beta}|$ from Eq.~\ref{eq:T12EffMass}.
Even though a strong effort is being put on the problem,
different nuclear models yield $\bigl| {\cal{M}}^{0\nu} \bigr|$ values which can vary
by more than a factor two. A compilation of possible estimations for the
most investigated \bb\ emitting isotopes is given in~\cite{Barea:2015kwa}.

The estimation of $g_A$ is still matter of debate~\cite{Barea:2013bz,Robertson:2013cy}.
Namely it is not clear if the value for free nucleons has to be considered,
or if some ``quenching'' is induced by limitations in the calculation
or by the omission of non-nucleonic degrees of freedom.
Given this ambiguity, the effect of the quenching of $g_A$ is not considered here,
and only unquenched values are used.

Among the dozen \bb\ decaying isotopes,
\Ge, \Te\ and \Xe\ are of particular interest due to the existence of established technologies
that led to the construction in the past years of several experiments
which set limits of $\sim0.2$--0.5~eV on
\meff~\cite{gerda:prl,Alfonso:2015wka,Albert:2014awa,Gando:2012zm}.
The next phase experiments employing these isotopes will allow to start probing
or partially cover the IH region, depending on the case.

\begin{table}
  \caption{\label{tab:PhaseSpaceNME}\onbb\ decay phase space factors,
    together with the smallest and largest NME
    for \Ge, \Te\ and \Xe.
    The values of $G^{0\nu}$ are taken from~\cite{Kotila:2012zza},
    the NME from~\cite{Barea:2015kwa}. For each NME, the nuclear model used
    for the calculation is reported. The value of $g_A$ considered for ISM
    is 1.25, for QRPA-T\"u $g_A=1.27$, and for IBM-2 $g_A=1.269$.}
  \hspace{-3mm}
  \begin{tabular}{rcccc}
    \hline\noalign{\smallskip}
            &  $G^{0\nu}$                          & \multicolumn{2}{c}{$\bigl| {\cal{M}}^{0\nu} \bigr|$} \\
    \cmidrule{3-4}
    Isotope & $[10^{\mbox{-}15}\text{yr}^{\mbox{-}1}]$ & Smallest & Largest \\
    \noalign{\smallskip}\hline\noalign{\smallskip}
    \Ge  & 2.363 & 2.81 (ISM)       & 5.16 (QRPA-T\"u) \\
    \Te & 14.22 & 2.65 (ISM)       & 3.89 (QRPA-T\"u) \\
    \Xe & 14.58 & 2.18 (QRPA-T\"u) & 3.05 (IBM-2) \\
    \noalign{\smallskip}\hline
  \end{tabular}
\end{table}

\begin{figure}
  \centering
  \def\svgwidth{\columnwidth}
  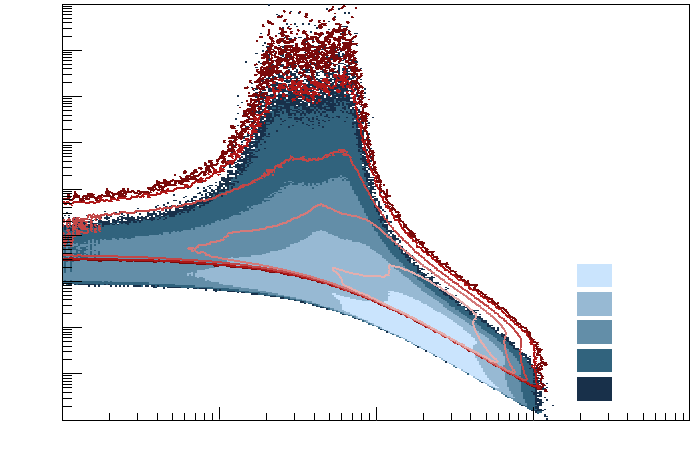
  \vspace{3mm}
  \def\svgwidth{\columnwidth}
  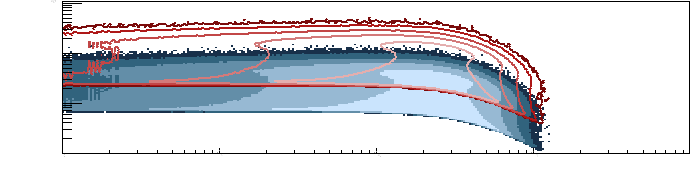
  \caption{\onbb\ decay half life for \Ge\ as a function of the lightest neutrino mass
    with the application of the cosmological bound.
    The blue regions are obtained with $|{\cal{M}}^{0\nu}| = 5.16$, the red curves with $|{\cal{M}}^{0\nu}| = 2.81$.
    The different color shadings correspond to the 1,\dots,5~$\sigma$ coverage regions.}
  \label{fig:WithC_Thalf_Mmin_Ge}
\end{figure}

\begin{figure}
  \centering
  \def\svgwidth{\columnwidth}
  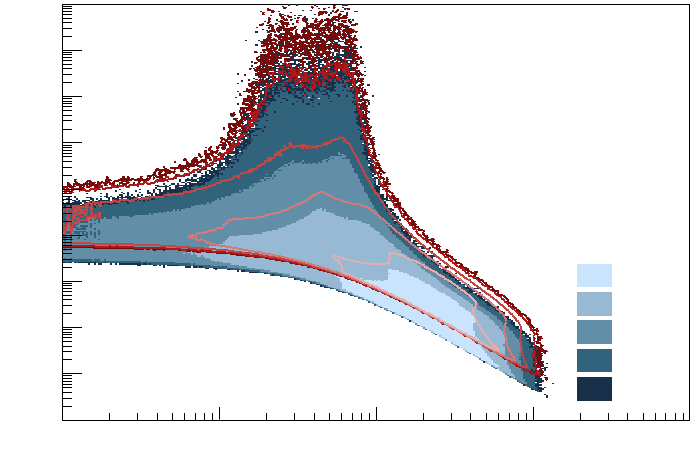
  \vspace{3mm}
  \def\svgwidth{\columnwidth}
  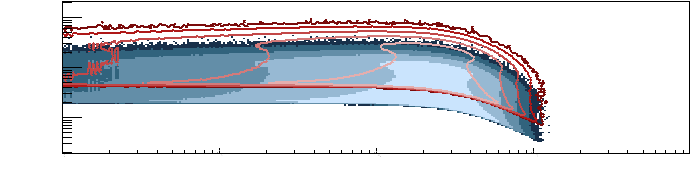
  \caption{\onbb\ decay half life for \Te\ as a function of the lightest neutrino mass
    with the application of the cosmological bound.
    The blue regions are obtained with $|{\cal{M}}^{0\nu}| = 3.89$, the red curves with $|{\cal{M}}^{0\nu}| = 2.65$.
    The different color shadings correspond to the 1,\dots,5~$\sigma$ coverage regions.}
  \label{fig:WithC_Thalf_Mmin_Te}
\end{figure}

\begin{figure}
  \centering
  \def\svgwidth{\columnwidth}
  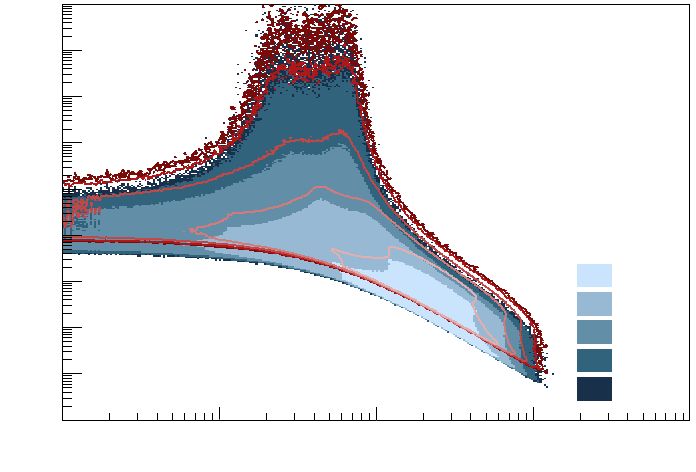
  \vspace{3mm}
  \def\svgwidth{\columnwidth}
  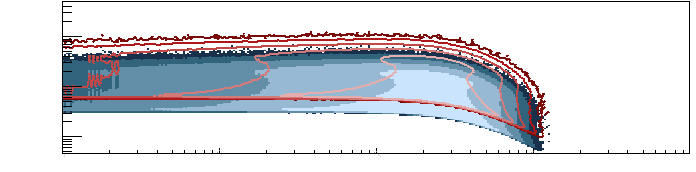
  \caption{\onbb\ decay half life for \Xe\ as a function of the lightest neutrino mass
    with the application of the cosmological bound.
    The blue regions are obtained with $|{\cal{M}}^{0\nu}| = 3.05$, the red curves with $|{\cal{M}}^{0\nu}| = 2.18$.
    The different color shadings correspond to the 1,\dots,5~$\sigma$ coverage regions.}
  \label{fig:WithC_Thalf_Mmin_Xe}
\end{figure}

From an experimental perspective, it is important to estimate the required sensitivity on \Tonbb\
in order to cover the IH or the NH region using \Ge, \Te\ and \Xe.
The probability distributions for \Tonbb\ as a function of \mmin\
with the application of the cosmological bound for the three isotopes
are shown in Figs.~\ref{fig:WithC_Thalf_Mmin_Ge}, \ref{fig:WithC_Thalf_Mmin_Te}
and~\ref{fig:WithC_Thalf_Mmin_Xe} for NH (top) and IH (bottom).
The contour lines correspond to the 1--5~$\sigma$ coverage regions.
The distributions have been extracted from that of \meff\
with the application of Eq.~\ref{eq:T12EffMass}.
The values for the phase space integral are taken from~\cite{Kotila:2012zza},
and the NME from~\cite{Barea:2015kwa},
and reported in Tab.~\ref{tab:PhaseSpaceNME}.
Given the large span between different NME estimations,
the largest and the smallest NME reported in~\cite{Barea:2015kwa} are used.
These correspond to the blue and red distributions in Figs~\ref{fig:WithC_Thalf_Mmin_Ge}, \ref{fig:WithC_Thalf_Mmin_Te}
and~\ref{fig:WithC_Thalf_Mmin_Xe}, respectively.
For \Ge, the IH band for the optimal NME starts at 6--7$\cdot10^{27}$~yr,
while for \Te\ and \Xe\ it starts at $2\cdot10^{26}$~yr and $3\cdot10^{26}$~yr, respectively.
In this regard, \Te\ and \Xe\ are preferable with respect to \Ge.
In reality, the total efficiency, the energy resolution, the atomic mass and the isotopic fraction
of the considered isotopes have to be considered, too.
A complete review of the topic is given in~\cite{GomezCadenas:2011it,Schwingenheuer:2012zs}.

If IH is assumed and considering the smallest NME,
an experiment aiming to measure \onbb\ decay needs a sensitivity on \Tonbb\
of $\sim4\cdot10^{28}$~yr if \Ge\ is employed, or of $\sim9\cdot10^{27}$~yr with \Te\ and \Xe.
Assuming NH and an effective mass of $10^{\mbox{-}3}$~eV, a \Ge\ based experiment
would need a sensitivity of $\sim6\cdot10^{30}$~yr, while with \Te\ and \Xe\
a $\sim10^{30}$~yr sensitivity is sufficient.

One remark can be made with respect to the debated Klapdor-Kleingrothaus claim of \onbb\ decay
observation in \Ge~\cite{KlapdorKleingrothaus:2004wj}.
In Fig.~\ref{fig:WithC_Thalf_Mmin_Ge} the $5~\sigma$ coverage region for the largest
NME (blue curves) does not extend below $10^{26}$~yr.
This is in very strong tension with the published $99.73\%$~CL interval for the \onbb\ decay half life,
$T_{1/2}^{0\nu}=(0.69-4.18)\cdot10^{25}$~yr~\cite{KlapdorKleingrothaus:2004wj}.
In other words, assuming that only the standard three light neutrino participate to \onbb\ decay
and using the largest NME and an unquenced $g_A$,
a $>5~\sigma$ disagreement is present between the cosmological bound
and the Klaptor-Kleingrothaus claim.

\section{Perspectives for Electron Neutrino Mass Measurements}
\label{sec:mbeta}

\begin{figure}
  \centering
  \def\svgwidth{\columnwidth}
  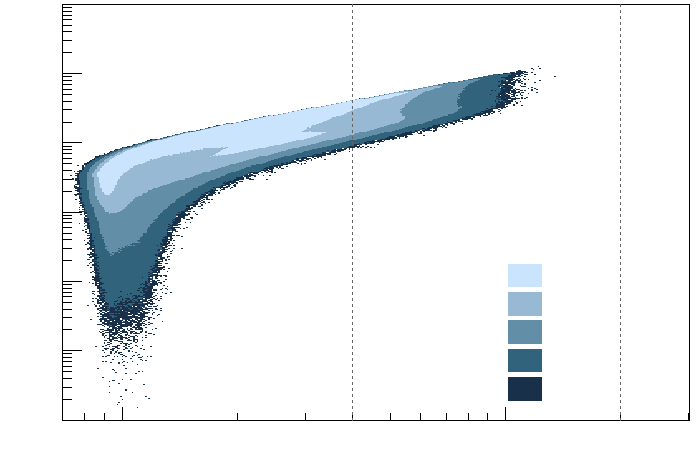
  \caption{Effective Majorana mass as a function of the electron neutrino mass
    for the NH regime with the application of the cosmological bound.
    The different color shadings correspond to the 1,\dots,5~$\sigma$ coverage regions.}
  \label{fig:WithC_Meff_Mbeta_NH}
\end{figure}

\begin{figure}
  \centering
  \def\svgwidth{\columnwidth}
  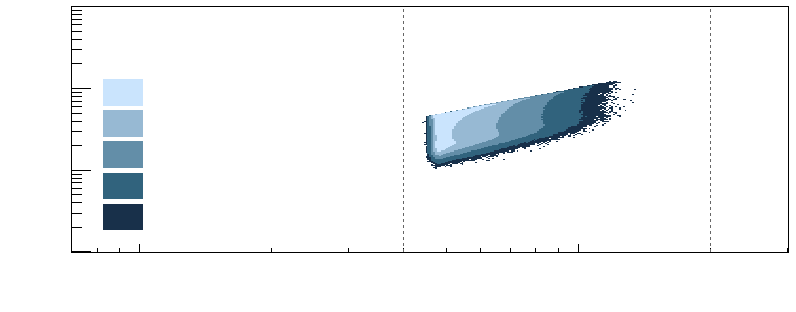
  \caption{Effective Majorana mass as a function of the electron neutrino mass
    for the IH regime with the application of the cosmological bound.
    The different color shadings correspond to the 1,\dots,5~$\sigma$ coverage regions.}
  \label{fig:WithC_Meff_Mbeta_IH}
\end{figure}

A side product of the present study is the behavior of \meff\
as a function of the electron neutrino mass, \mbeta.
This is the parameter of interest for the experiments measuring
the end point of beta spectra, e.g. KATRIN~\cite{Osipowicz:2001sq} and Project8~\cite{Doe:2013jfe},
or the $^{163}$Ho electron capture, e.g. ECHo~\cite{Gastaldo:2013wha} and HOLMES~\cite{Alpert:2014lfa} .
The electron neutrino mass is given by:
\begin{align}\label{eq:mbeta}
  m_{\beta} & = \sqrt{ \sum_{i=1}^3 |U_{ei}|^2 m_i^2 } \nonumber \\
  & = \sqrt{ c_{13}^2 c_{12}^2 m_1^2 + c_{13}^2 s_{12}^2 m_2^2 + s_{13}^2 m_3^2 }
\end{align}
where no new physics is involved with respect to the standard model.
Figs.~\ref{fig:WithC_Meff_Mbeta_NH} and~\ref{fig:WithC_Meff_Mbeta_IH}
show \meff\ as a function of \mbeta\ for the NH and IH, respectively.
The minimum of \mbeta\ comes from the values of $\delta m_{\sun}^2$
and $\Delta m_{atm}^2$, while the shading for $m_{\beta}\gtrsim5\cdot10^{\mbox{-}2}$~eV
is governed by the cosmological limit on $\Sigma$.
The vertical dashed lines correspond to the sensitivity of KATRIN at $0.2$~eV~\cite{Osipowicz:2001sq,Drexlin:2004as},
and of the recently proposed Project8 experiment at 40~meV~\cite{Doe:2013jfe}.
It is interesting to see how strongly the cosmological observations
affect the expectation for \mbeta.
Assuming $\Sigma = (22\pm62)\cdot10^{\mbox{-}3}$~eV~\cite{Palanque-Delabrouille:2014jca,Dell'Oro:2015tia},
the probability for KATRIN to find a signal is practically zero.
Even using an arbitrarily looser bound on $\Sigma$,
the chance for \mbeta\ to be at the 0.2~eV level remains very limited.
On the contrary, the cosmological bound does not really affect the physics reach of Project8.
Its goal sensitivity is sufficient to fully cover the IH region:
if nature chose IH, the cosmological bound cannot exclude all values of \mbeta\ down to 40~meV
without being in conflict with the measurements of $\delta m_{\sun}^2$ and $\Delta m_{atm}^2$.
If the target sensitivity is reached, Project8 would have very good chances of discovery.
If neutrino masses follow the NH scheme,
the allowed range for \mbeta\ would extend down to $\sim8$~meV.
At present, no technology is capable of reaching such a sensitivity,
and the possibility to measure the electron neutrino mass might be unreachable.

\section{Conclusions}

In this work, a new method for the calculation of the allowed range for \meff\
and \Tonbb\ in the standard three neutrino scheme is proposed.
It is based on the random sampling of the involved mixing angles,
of the squared neutrino masses and of the Majorana phases.
The effect of these on \meff\ is highlighted,
and the consequences for \onbb\ decay search are described.
The assumption of a flat distribution in the $[0,2\pi]$ region for the Majorana phases
yields a $\geq93\%$ discovery probability for an experiment with $10^{\mbox{-}3}$~eV sensitivity on \meff.
Smaller \meff\ values can be obtained only if the Majorana phases differ by $\sim\pi$.
Based on this, theoretical models for the neutrino mass matrix predicting
a small \meff~\cite{Vissani:2003aj,Vissani:2001im} are expected to have Majorana phases which obey this condition.

In a second step, the cosmological bound on $\Sigma$ is applied,
leading to the two dimensional probability distribution for \meff\
as a function of the lightest neutrino mass.
A weak preference for \meff\ values close to the degenerate region is found,
although the strong dependence on the probability distribution for $\Sigma$
invokes some caution in the extraction of predictions.
A $3.3\cdot10^{\mbox{-}3}$~eV discovery sensitivity on \meff\
in case of NH and $2.1\cdot10^{\mbox{-}2}$~eV in case of IH
are required to future experiments in order to achieve
a $90\%$ probability of measuring \onbb\ decay.

Moreover, the probability distribution for the \onbb\ decay half life
as a function of the lightest neutrino mass for \Ge, \Te\ and \Xe\ is given.
This shows how a sensitivity of $O(10^{28})$~yr on \Tonbb\ is required
for all three isotopes to cover the IH region,
while in case of NH a sensitivity of $O(10^{30})$~yr is needed.

Finally, the perspectives for the direct measurement of the electron neutrino mass
with $\beta$-decay end point experiments are discussed,
and the possibility of discovery for an experiment with 40~meV sensitivity is shown,
under the assumption that neutrino masses are distributed according to the IH.

\begin{acknowledgements}
The author gratefully thanks L.~Baudis, R.~Brugnera, G.~Isidori, L.~Pandola, B.~Schwingenheuer and F.~Vissani
for the enlightening suggestions, and P.~Grabmayr and B.~Schwingenheuer for carefully checking the manu-script.
\end{acknowledgements}

\bibliographystyle{h-physrev5}
\bibliography{Bibliography}

\end{document}